\documentclass[twocolumn]{article}
\pdfoutput=1

\usepackage{nicefrac}
\usepackage{subfigure}
\usepackage{geometry}
\usepackage{graphicx}
\usepackage{amssymb,amsmath,siunitx}
\usepackage[infoshow,debugshow]{tabularx}
\usepackage{color}
\usepackage{cleveref}
\usepackage{indentfirst}
\usepackage{setspace}
\usepackage[british]{babel}
\usepackage{placeins}
\usepackage{authblk}

\DeclareMathOperator{\diver}{div}
\DeclareMathOperator{\diag}{diag}
\DeclareMathOperator{\tr}{tr}

\renewcommand{\d}{\operatorname{d}}

\newcommand{\crt}{\operatorname{cr}}
\newcommand{\T}{\operatorname{T}}

\begin{document}

\title{The mechanics of human brain organoids}

\author[1]{Valentina Balbi}
\author[1]{Michel Destrade}
\author[2]{Alain Goriely}

\affil[1]{\small{School of Mathematics, Statistics and Applied Mathematics, NUI Galway,  University Road, Galway, Ireland}}
\affil[2]{Mathematical Institute, University of Oxford}

\date{}

%%%%%%%%%%%%%

\twocolumn[
  \begin{@twocolumnfalse}
    \maketitle
   \begin{abstract}
   
Organoids are prototypes of human organs derived from cultured human stem cells. 
They provide a reliable and accurate experimental model to study the physical mechanisms underlying the early developmental stages of human organs  morphogenesis and, in particular, the early morphogenesis of the cortex.
%However, human brain organoids are particularly difficult to manufacture, given the complex nature and structure of the organ. 
Here, we propose a mathematical model to elucidate the role played by two mechanisms which have been experimentally proven to be crucial in shaping human brain organoids: the contraction of the inner core of the organoid and the microstructural remodeling of the outer cortex. 
Our results show that both mechanisms are crucial for the final shape of the organoid and can explain the origin of brain pathologies such as lissencephaly (smooth brain).
\vspace{2em}	
	\end{abstract}

  \end{@twocolumnfalse}
]

\maketitle

%%%%%%%%%%%%%%%

\begin{figure}[th]
\includegraphics[width=8cm]{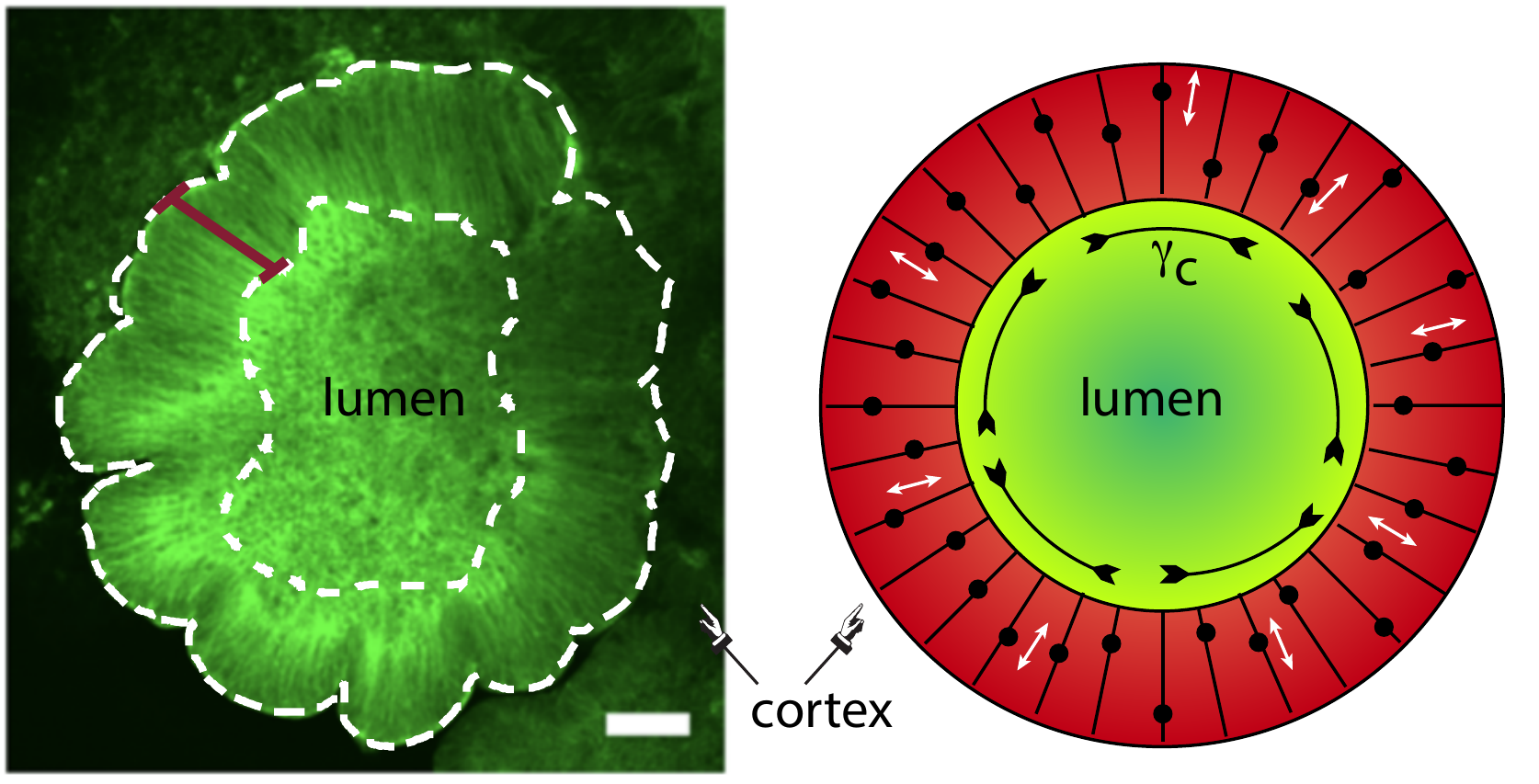}
\caption{Left: Morphology of a growing organoid (adapted from \cite{karzbrun2018human}, scale bar is 100$\mu$m) and (Right) a two-dimensional morphoelastic model. The black circles represent the nuclei that move up and down radially while the lumen can contract in the hoop direction.}\label{Fig1}
\end{figure}

\indent\textbf{Introduction --\ }The characteristic convoluted shape of the human brain was first reported in the Edwin Smith papyrus, an Egyptian manuscript dated 1,700BC that compares brain convolutions to the corrugations or wrinkles found in molten metal \cite{bearsted1930edwin}. The description, development, and function  of these convolutions have also been major topics of research for the last two centuries \cite{gobuku15}. %Since then the brain has fascinated scientists for its peculiar and beautiful geometry. 
The visible upper part of these convolutions are called \textit{gyri}  and their deep groves are referred to as \textit{sulci}. Geometrically, the convolutions increase the surface area of the brain for a given volume. From a functional point of view, it is believed that they have the strategic functions of increasing the number of neuronal bodies located in the cortex and facilitating the connections between neurons hence reducing the traveling time of the electric signals between different regions. 
Although different explanations have been proposed, the mechanisms behind \textit{gyrification} are  not fully understood. 
It is now accepted that intrinsic mechanical forces, rather than external constraints, are responsible for the emergence of folding in the human brain \cite{striedter2015cortical} and recent observational studies \cite{ronan2013differential,garcia2018dynamic} further support the role of the rapid tangential expansion of the cortex during development as the primary driver for folding \cite{richman1975mechanical,bayly2014mechanical,budday2014role,bayly2013cortical,gobuku15}. 
%Early studies focused on the hypothesis that the axons, distributed within the white matter core, pull the neuronal cells, located in the cortex, inwardly and eventually induce the folded patterns found on the surface of the cortex \cite{van1997tension}. 
%But it has been  shown experimentally that although the cortex is under radial tension, this force is not strong enough to induce buckling in the cortical layer \cite{xu2010axons}. 
%Another current of thought supports the hypothesis of differential growth between the white matter core and the grey matter cortex \cite{tallinen2014gyrification,budday2014role,bayly2013cortical}. 
At the simplest physical level, the onset of folding can be understood as an initial build up of elastic energy in the compressed upper cortex and its partial release by a wrinkling deformation of  the film and substrate.  Experimentally, this instability can be observed  in  the constrained polymeric swelling of a circular shell  bounded to an elastic disk which triggers the same type of wrinkling pattern \cite{mogo11,dervaux2011buckling,ciarletta2014pattern,balbi2015morphoelastic}. Similar experiments performed on a two-layered brain prototype made of polymeric gels  with differential swelling properties reproduce folds similar to the gyri and sulci of a real brain \cite{tallinen2016growth}. \\
\indent 
%Experimentally, different techniques have been used to investigate brain development and brain wrinkling. 
%Tallinen et al. \cite{tallinen2016growth} 3D-printed a two-layered brain prototype made of polymeric gels, based on the size and shape of a 22-week foetal brain. 
%Immersion in a nutrient solution induced swelling in the outer layer of the brain prototype, which in turn initiated folds similar to the gyri and sulci of a real brain. 
Yet, the brain is too complex to be used as an experimental platform to unravel the detailed cellular mechanisms leading to folding. An alternative approach is to use \textit{human brain organoids}. Organoids are self-organized and collective structures produced in vitro from the culturing of human stem cells that mimic the  early development of the human brain. In particular, when cultured in a mostly flat geometry, these organoids develop  wrinkling patterns as shown recently by
Karzbrun et al. \cite{karzbrun2018human} (Fig.~\ref{Fig1}).
 Moreover, these authors  identified  two possible mechanisms  for surface folding during cortical development: the contraction of the organoid lumen and the nuclei motion of neuronal progenitor cells within the cortical layer. Different morphologies can be obtained by varying these effects. The key problem is then to develop a physical model to understand the relative role of each effect in creating folding patterns in brain organoids. Here, we use morphoelasticity \cite{goriely17}, the theory of growing and remodeling continua, to model brain organoids and explore various parameter regimes.\\

\indent\textbf{The model --\ }  The organoid is modeled as a $2$D continuum morphoelastic structure  made of an inner disc, the lumen, which serves as the core of the organoid, and a surrounding ring, the cortex. 
%First, we describe the kinematics of the problem.
%Next, we write down and solve the elastic equilibrium problem which fully describes the organoid dimensions and internal state of stress after contraction of the lumen and microstructural remodeling of the cortical layer. 
%Finally, we perturb this state to study the emergence of instability wrinkling patterns and validate the predictions of our model with the experimental data available in the literature.\\
%
We introduce a cylindrical coordinate system so that a material point   at position  $\textbf{X}\in\mathbb{R}^{2}$ with components $(R,\Theta)$ in the  initial configuration $\mathcal{B}_0$ is at position $\textbf{x}\in\mathbb{R}^{2}$ with components $(r,\theta)$ in the final, deformed, configuration $\mathcal{B}$. The deformation $\textbf{x}=\textbf{x}(\textbf{X}) $ from $\mathcal{B}_0$ to $\mathcal{B}$ defines the  associated deformation gradient $\textbf{F} = \partial \textbf x/\partial \textbf{X}$.
%
%We call $\boldsymbol{\chi}:\mathcal{B}_0\rightarrow\mathcal{B}$ the deformation that transforms $\textbf{X}$ into $\textbf{x}$ and  the\
%
\indent 
To model the contraction of the lumen (shrinking) and the remodeling of the cortex, we use the multiplicative decomposition of the deformation gradient.
Hence, $\textbf{F}$ can be split into a tensor $\textbf{G}$  describing the natural growth or remodeling of the tissue and introducing a virtual intermediate incompatible state; and an elastic part $\textbf{A}$ restoring the compatibility of the structure, whilst possibly introducing residual stresses \cite{goriely17}. 
Specifically, the lumen undergoes a uniform contraction that results in an isotropic shrinking in the plane, whilst microstructural changes occur at constant volume within the cortex due to the radial motion of cell bodies. The tensor $\textbf{G}$ for the lumen and the cortex is:
\begin{equation}
\textbf{G}_L=\diag(\gamma_L,\gamma_L)\,\,\text{and}\,\,\textbf{G}_C=\diag(\gamma_C,1/\gamma_C)
\end{equation}
where $\gamma_L<1$ is a measure of the volumetric contraction of the lumen and $\gamma_C$ is the remodeling stretch associated with microstructural changes in the cortex. 
\begin{figure}[t]
\includegraphics[width=8cm]{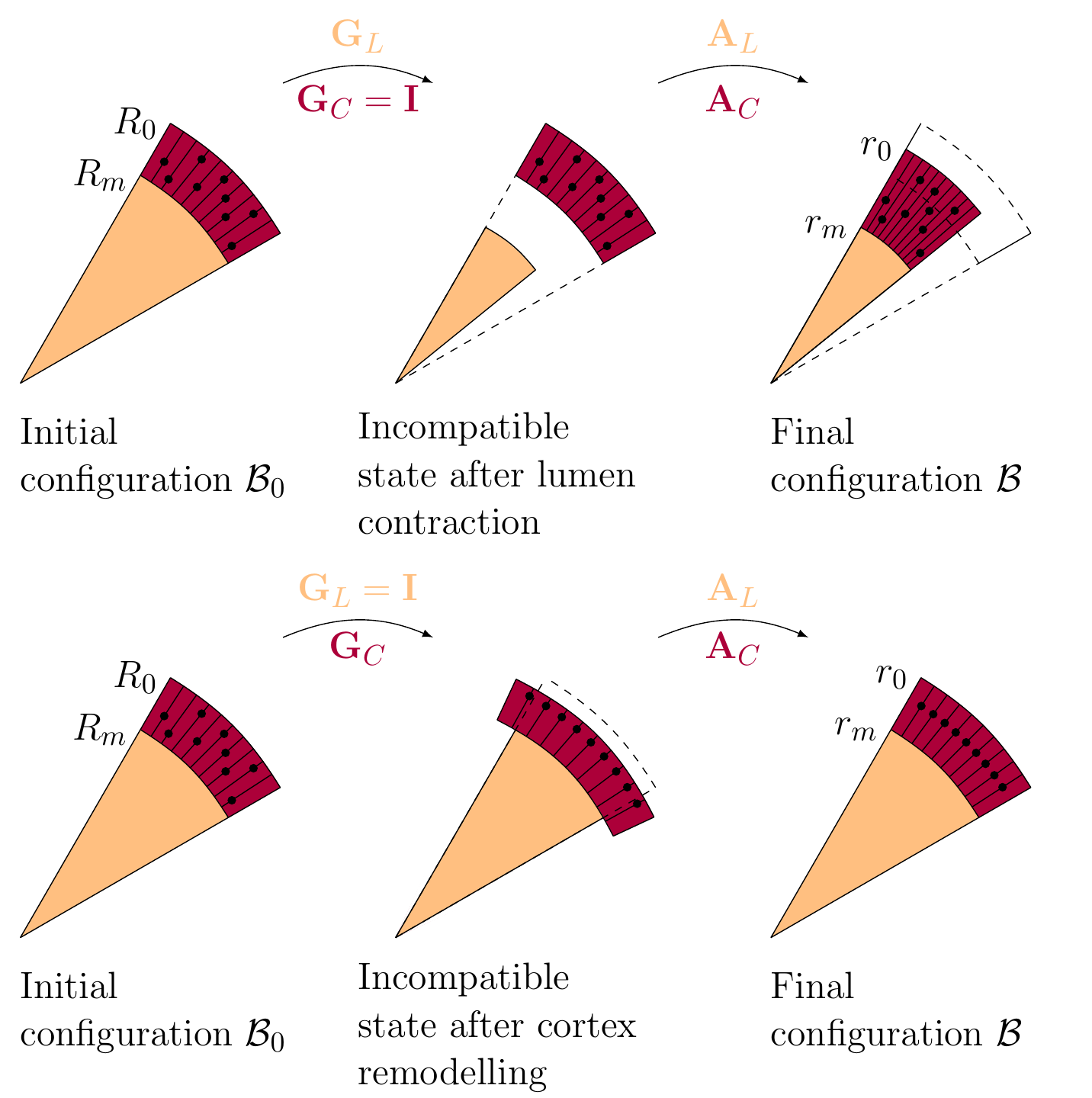}
%\subfigure[]{\includegraphics[scale=0.53]{Fig2a.pdf}}
%\subfigure[]{\includegraphics[scale=0.53]{Fig2b.pdf}}
\caption{Schematic illustration of (a) the contraction of the lumen and (b) the remodeling of the cortex (small dots represent nuclei).}\label{fig:contr_rem}
\end{figure}
%Furthermore, the cortex remodels at constant volume, which explains that $\det \textbf{G}_C=1$ at all times. 
Fig.~\ref{fig:contr_rem}(a) illustrates the deformation resulting from the contraction of the lumen and Fig.~\ref{fig:contr_rem}(b) shows how a slice of the organoid deforms when remodeling occurs in the cortex.
This form of $\textbf G_C$ captures  the observed back and forth motion of the nuclei of  neuron progenitor cells, because it  generates inwards or outwards forces within the cortex, as we show later. \\
\indent
Both the lumen and the cortex are assumed incompressible and a simple base solution for the deformation maintaining the circular symmetry is:
\begin{equation}\label{axialsymm}
\begin{split}
r&=\gamma_L R,\qquad \quad \theta=\Theta \quad \text{for}\quad R\!\in\![0,R_m)\\
r&=\sqrt{R^2+a},\quad \theta=\Theta\quad \text{for}\quad R\!\in\![R_m,R_0]
\end{split}
\end{equation} 
where $a\!=\!r_m^2\!-\!R_m^2$ is the unique unknown parameter (to be specified from the mechanical equilibrium), and $R_m,R_0$ ($r_m, r_0$) are the interface and outer radii in the undeformed (deformed) configuration. 
The initial geometry of the structure is  defined by the non-dimensional aspect ratio $H\!=\!R_0/R_m$.
Both the lumen and the cortex are assumed to follow a hyperelastic, isotropic and neo-Hookean behavior so that for each tissue, the existence of a strain energy function in the form $W_N\!=\!\mu_N(I_{1N}-3)/2$ with $N\!=\!\{L,C\}$ can be postulated. 
Here, $\mu_L$ and $\mu_C$ are the shear moduli of the lumen and cortex, respectively, and define the non-dimensional stiffness ratio $\mu= \mu_C/\mu_L$, and $I_{1N}=\tr\textbf{B}_N$ is the first invariant of the left Cauchy-Green tensor $\textbf{B}_N\!=\!\textbf{A}_N\textbf{A}_N^{\T}$. 
The constitutive equation for the Cauchy stress can then be written down for each tissue as $\boldsymbol{\sigma}_N\!=\!\mu_N\textbf{B}_N-p_N\textbf{I}$, $N\!=\!\{L,C\}$, where $p_N$ are the Lagrange multipliers introduced to enforce incompressibility and $\textbf{I}$ is  the $2 \times 2$ identity tensor. 
\\
\indent 
The elastic equilibrium problem is given by:
\begin{equation}\label{eqprob}
\dfrac{\text{d}}{\text{d} r}\sigma_{Nrr}(r)+\dfrac{\sigma_{Nrr}(r)-\sigma_{N\theta\theta}(r)}{r}=0,\quad N=\{L,C\}
\end{equation}
with the boundary and the continuity conditions 
\begin{equation}\label{bvp}
\sigma_{Crr}(r_0)=0,\quad \text{and}\quad\sigma_{Lrr}(r_m)=\sigma_{Crr}(r_m).
\end{equation}
The solution to \eqref{eqprob} with \eqref{bvp}  is:
\begin{equation}\label{eqsol}
\begin{split}
\sigma_{Crr}(r)&=\int_r^{r_0}\mu\dfrac{B_{Crr}(r)-B_{C\theta\theta}(r)}{r}\d r,\\
\sigma_{C\theta\theta}(r)&=\mu\left(B_{C\theta\theta}(r)-B_{Crr}(r)\right)+\sigma_{Crr}(r),\\
\sigma_{Lrr}(r)&=\sigma_{Lrr}(r_m),\qquad\sigma_{L\theta\theta}(r)=\sigma_{L\theta\theta}(r_m).
\end{split}
\end{equation}
The solution \eqref{eqsol} (details given in the Supplementary Material% at [URL will be inserted by publisher]
) shows that  the Cauchy stress is constant within the lumen because the deformation is homogeneous there. The hoop stress is negative within the cortex, indicating that the outer layer is under circumferential compression.  We  expect that for some critical values of the parameters,  the organoid will buckle to release the  build-up of compressive residual stress in the cortex. Further, when $\gamma_C\!<\!1$ a positive radial stress is generated and the force gradient favors the outward nuclei motion. When $\gamma_C\!>\!1$ the radial stress component is negative and contributes to pushing the nuclei inwardly, according to the observed cyclic outward/inward nuclear motion within the cortex.\\
Next we study  the stability of the symmetric solution in \eqref{axialsymm} and \eqref{eqsol}. We perturb the current position vector $\textbf{x}$ with components in \eqref{axialsymm}, by superposing an incremental \cite{bego05} deformation $\delta\textbf{x}$. 
% with components:
%\begin{equation}
%u(r,\theta)=U(r)\cos(m\theta),\quad v(r,\theta)=V(r)\sin(m\theta).
%\end{equation}
The  perturbed position vector is $\tilde{\textbf{x}}_N\!=\!\textbf{x}_N+\varepsilon\delta\textbf{x}_N$ ($N\!=\!\{L,C\}$) and the associated elastic deformation gradient is $\tilde{\textbf{A}}_N\!=\textbf{A}_N\!+\!\varepsilon\boldsymbol{\Gamma}_N\textbf{A}_N$, where $\boldsymbol{\Gamma}_N\!=\!{\partial\delta\textbf{x}_N}/ {\partial\textbf{x}_N}$ is the incremental displacement gradient. Accordingly, we define the perturbed stress $\tilde{\boldsymbol{\sigma}}_N\!=\!\boldsymbol{\sigma}_N+\varepsilon\delta\boldsymbol{\sigma}_N$ where:
\begin{equation}\label{incrstress}
\delta\boldsymbol{\sigma}_N\!=\!\boldsymbol{\mathcal{A}}_N\!:\!\boldsymbol{\Gamma}_N^{\T}\!+p_N\delta\textbf{F}_N-\delta p_N\textbf{I}
\end{equation} is the push-forward of the incremental nominal stress. 
Here $\boldsymbol{\mathcal{A}}_N \!=\!\textbf{A}_N( \partial^2 W_N / \partial \textbf{A}_N^2) \textbf{A}_N$ is the fourth-order tensor of the instantaneous elastic moduli and $\delta p_N$ is the increment of the Lagrange multiplier.  Note that $\boldsymbol{\mathcal{A}}_L$ is constant because the deformation is homogenous within the lumen. The incremental problem then amounts to solving:
\begin{equation}\label{increq}
\diver\delta \boldsymbol{\sigma}_N=\boldsymbol{0} \quad\text{and}\quad\tr\boldsymbol{\Gamma}_N\!=\!0,\qquad N=\{L,C\}
\end{equation} 
the latter being the incremental incompressibility condition. \\
\indent 
We seek a wrinkling solution of the form:
\begin{equation}\label{incrsol}
\begin{split}
\{\delta r_N,\delta p_N\}&=\{U_N(r),P_N(r)\}\cos(m\,\theta)\\
\delta \theta_N&=V_N(r)\sin(m\,\theta)
\end{split}
\end{equation}
where $\delta r_N,\delta\theta_N$ are the components of the incremental displacement $\delta\textbf{x}_N$ and $m$ is the wave-number of the perturbation. 
To tackle the incremental problem \eqref{increq}, which is a second-order linear system of three partial differential equations with boundary conditions, we transform it into a first-order linear system of four ordinary differential equations (ODEs) with initial conditions. 
From \eqref{increq} and \eqref{incrsol} we find that the incremental stress components have the form: $(\delta\boldsymbol{\sigma}_N)_{rr}\!=\!\Sigma_{Nrr}(r)\cos(m\,\theta)$ and $(\delta\boldsymbol{\sigma}_N)_{r\theta}\!=\!\Sigma_{Nr\theta}(r)\sin(m\,\theta)$ and we re-write \eqref{increq} as
\begin{equation}\label{stroh}
\dfrac{\d}{\d r}\boldsymbol{\eta}_N(r)=\dfrac{1}{r}\textbf{M}_N(r)\boldsymbol{\eta}_N(r),\qquad N=\{L,C\}
\end{equation}
where $\boldsymbol{\eta}_N(r)\!=\!\{\textbf{U}_N(r),r\boldsymbol{\Sigma}_N(r)\}^{\T}$ and the $4 \times 4$  $\textbf{M}_N$ is the so-called Stroh matrix \cite{govade08,destrade2001explicit,destrade2009,norris2010wave}. 
Note that \eqref{stroh} is a system of ODEs with variable coefficients. 
In the lumen, since  $\textbf{M}_L$ has constant components
 we have:
 \begin{equation}\label{strohsolL}
\boldsymbol{\eta}_L(r)=c_1 r^{m-1}\boldsymbol{\eta}^{(1)}(r)+c_2 r^{m+1} \boldsymbol{\eta}^{(2)}(r),
\end{equation}
where $m\!-\!1$ and $m\!+\!1$ are the eigenvalues of $\textbf{M}_L$ and $\boldsymbol{\eta}^{(i)}(r)\!=\!\{\boldsymbol{\eta}_{\textbf{U}i}(r),\boldsymbol{\eta}_{\boldsymbol{\Sigma}i}(r)\}^{\T}$ with $i\!=\!\{1,2\}$ are the associated eigenvectors. \\
\indent 
To  solve numerically the Stroh problem in the cortex  (method given in Supplementary Material% at [URL will be inserted by publisher]
), we fix $H$ and $\mu$, and by iterating over the wavenumber $m$ and either $\gamma_C$ or $\gamma_L$, we solve for  $\boldsymbol{\eta}$  from $r_0$ to $r_m$  with null initial condition until the outer boundary  conditions are satisfied.\\

\begin{figure}[t]
\includegraphics[width=8cm]{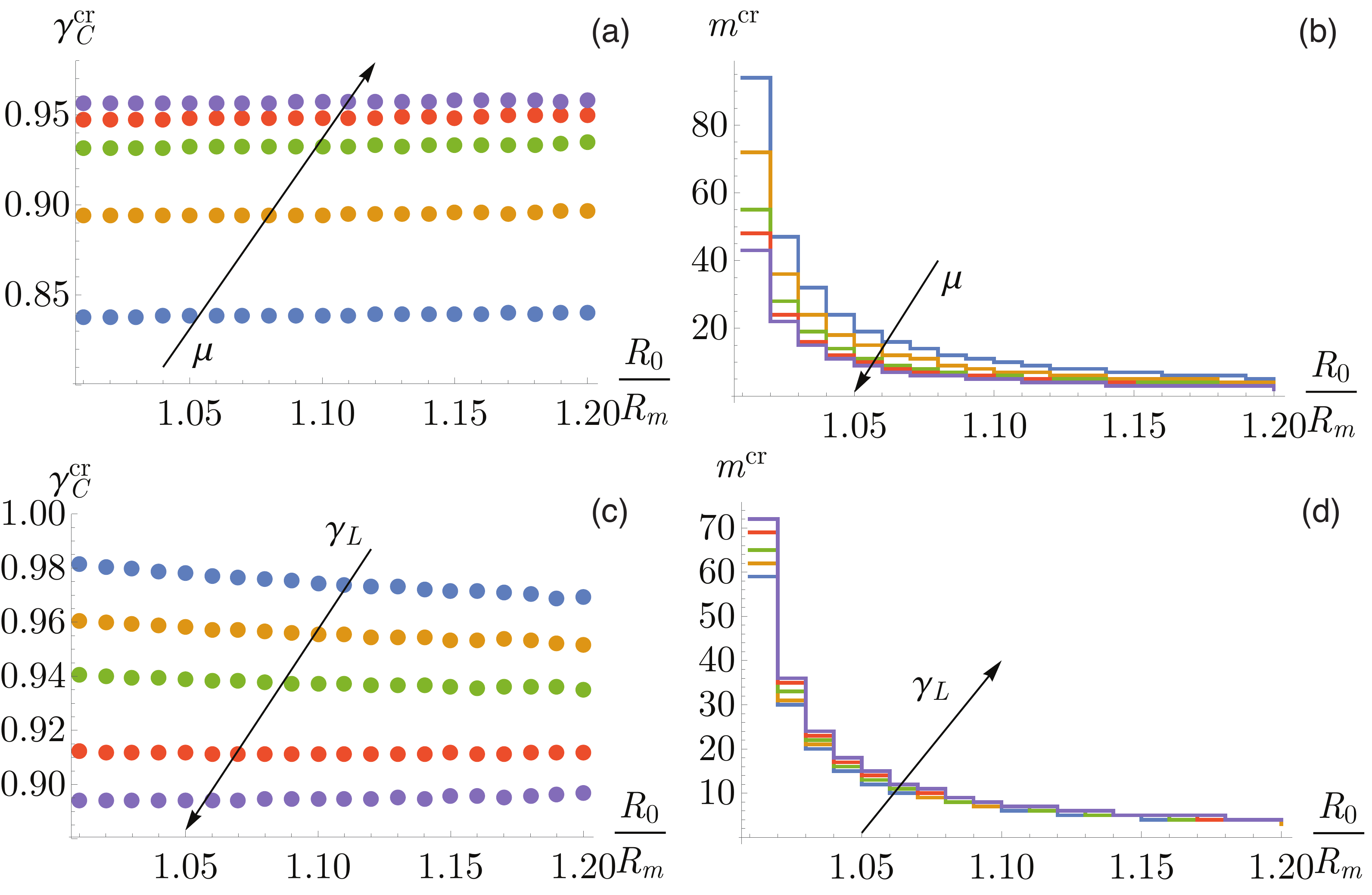}
%\subfigure[][]{\includegraphics[scale=0.238]{Fig3a}}
%\subfigure[][]{\includegraphics[scale=0.238]{Fig3b}}
%\subfigure[][]{\includegraphics[scale=0.238]{Fig3c}}
%\subfigure[][]{\includegraphics[scale=0.238]{Fig3d}}
\caption{Critical instability thresholds $\gamma_C^{\crt}$ and wavenumbers $m^{\crt}$ against the initial aspect ratio $H\!=\!{R_0}/{R_m}$ at varying stiffness ratios $\mu\!=\! {\mu_C}/{\mu_L}\!=\!\{5,10,20,30,40\}$ with fixed $\gamma_L=1$, (a) and (b); and at varying $\gamma_L\!=\!\{0.91,0.93,0.95,0.98,1\}$ with fixed $\mu=10$, (c) and (d).}\label{fig:resstab}
\end{figure}

\indent\textbf{Results --\ }Fig.~\ref{fig:resstab} shows the critical instability thresholds $\gamma_C^{\crt}$ against the initial aspect ratio ${R_0}/{R_m}$ at varying $\gamma_L$ and $\mu$ with the associated critical wavenumbers $m^{\crt}$.
From Figures \ref{fig:resstab} (a) and (b) we  conclude that the stiffer the cortex, the higher the critical remodeling thresholds $\gamma_C^{\crt}$. Therefore, disks with a stiff ring will buckle earlier (in the sense that the remodeling in the cortex  introduces less circumferential compression to trigger the instability) than those with a soft ring. These results are in accordance with those found for a substrate with a growing  layer \cite{bustgo15}. Another interesting aspect arises from Fig.~\ref{fig:resstab} (a): when the lumen does not contract, the cortex instability thresholds $\gamma_C^{\crt}$ are independent of the initial thickness of the cortical ring, i.e. the aspect ratio $H$. 
However, the instability patterns ($m^{\crt}$ in Fig.~\ref{fig:resstab} (b)) do depend on the thickness of the cortex (particularly for very thin rings). 
Finally, as the lumen contracts (upper curves in (c), $\gamma_L<1$), the critical thresholds $\gamma_C^{\crt}$ for the cortex decrease, indicating that the more the lumen contracts, the earlier the cortex will buckle. 
This is expected because the lumen contraction introduces a compressive circumferential stress in the cortex. 
%Moreover, increasing the stiffness ratio $\mu$, i.e. reducing the stiffness of the lumen $\mu_L$, has the same effect on the instability pattern, as inhibiting the lumen contractility. 
\begin{figure}[t]
\includegraphics[width=8cm]{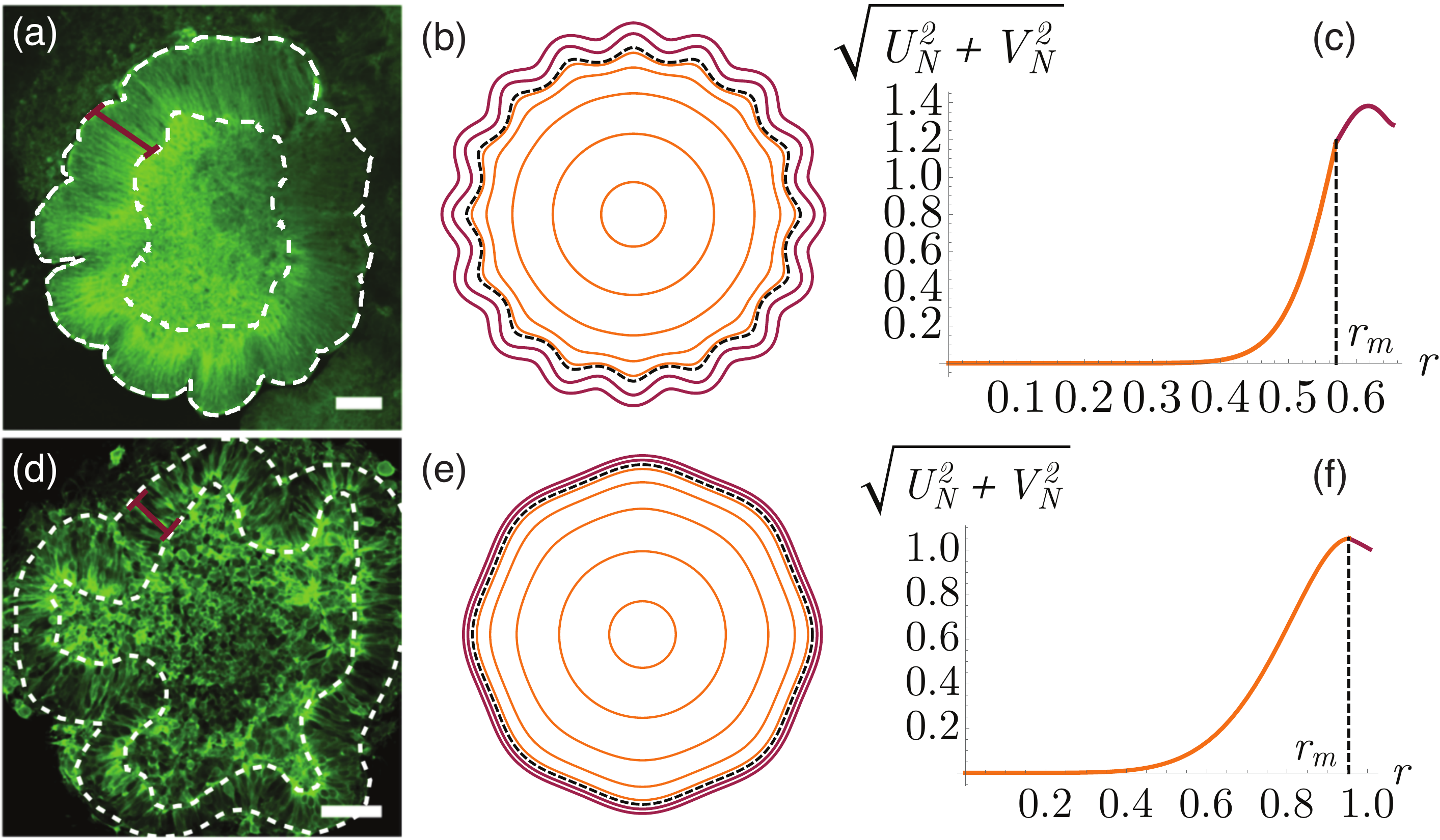}
%\subfigure[][]{\includegraphics[scale=0.25]{Fig4a}}
%\subfigure[][]{\includegraphics[scale=0.15]{Fig4b}}
%\subfigure[][]{\includegraphics[scale=0.15]{Fig4c}}
%\subfigure[][]{\includegraphics[scale=0.25]{Fig4d}}
%\subfigure[][]{\includegraphics[scale=0.15]{Fig4e}}
%\subfigure[][]{\includegraphics[scale=0.15]{Fig4f}}
\caption{Effect of contraction inhibition on the final morphology of the organoid: (a) and (d), control and drug-treated organoids (digitized from \cite{karzbrun2018human}, scale bars are 100$\mu$m); (b) and (e), instability patterns predicted by our model from an initial organoid with $H\!=\!1.05$, $\gamma_C\!=\!1$ and by setting $\mu\!=\!0.85$
% $\gamma_L^{\crt}\!=\!0.57$, $m^{\crt}\!=\!16$ 
 in (b), and $\mu\!=\!40$
%$\gamma_L^{\crt}\!=\!0.95$, $m^{\crt}\!=\!8$ 
 in (e); (c) and (f), magnitude of the incremental displacements $U_N(r),V_N(r)$, $N=\{L,C\}$ within the organoid (orange and purple, respectively), computed using the same parameters as in (b) and (e), and taking $\varepsilon\!=\!0.02$.}\label{fig:contr}
\end{figure}\\
\indent 
In Fig.~\ref{fig:contr}, we show the final morphologies obtained from two disks with the same initial aspect ratio $H\!=\!1.05$, but with different stiffness ratios $\mu$. 
In (e) the lumen is softer than the cortex ($\mu\!=\!40$) and the opposite scenario is modeled in (b) ($\mu\!=\!0.85$). The critical threshold for the instability in (e) $\gamma_L^{\crt}\!=\!0.95$ is much higher than that in (b) $\gamma_L^{\crt}\!=\!0.57$, indicating that disks with softer cores buckle at lower contraction thresholds. 
\\

\begin{figure}[t]
\centering
\includegraphics[width=8cm]{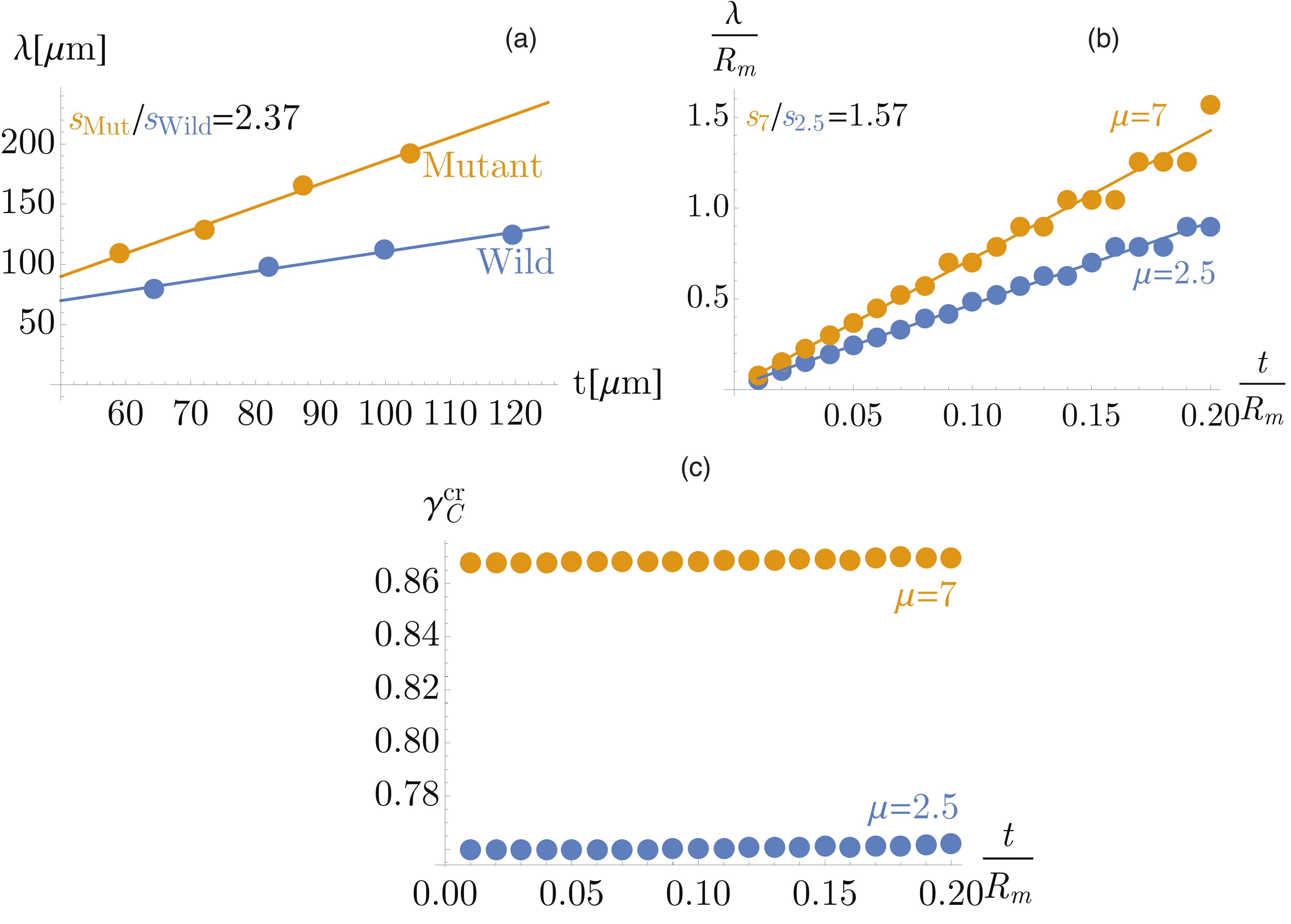}
%\subfigure[]{\includegraphics[scale=0.19]{Fig5a}}
%\subfigure[]{\includegraphics[scale=0.19]{Fig5b}}\\
%\subfigure[]{\includegraphics[scale=0.2]{Fig5c}}
\caption{LIS1 mutation: (a) Wavelength $\lambda$ vs thickness $t$. Data and linear fit for control (blue) and for mutant organoids (orange) (adapted from \cite{karzbrun2018human}), with the slope ratio $s_{\text{Mut}}/s_{\text{Wild}}$; (b) linear relation between the wavelength and the cortex thickness predicted by the presented model ($\gamma_L=1$), with the slope ratio $\mu_{7}/\mu_{2.5}$; (c) corresponding critical thresholds $\gamma_C^{\crt}$.}\label{fig:LIS1}
\end{figure}

\indent\textbf{Discussion --\ }These results  capture the effect of contraction inhibition of the cytoskeleton in human organoids. 
Indeed, in \cite{karzbrun2018human} organoids treated with Blebbistatin (a drug used to reduce the contractility of the lumen) were found to have a softer core (Case (e)) than control organoids (Case (b)). 
Moreover, our model predicts that the critical wavenumber for a disk with a soft core, $m^{\crt}=8$, is lower than that of a disk with a stiffer core, $m^{\crt}=16$. 
In fact, the treated organoids display less folds than control ones (compare Figures \ref{fig:contr} (a,b) with (d,e)). 
\\
\indent 
Furthermore, by comparing the magnitude of the instability patterns in Figures \ref{fig:contr} (c) and (f) we can observe that in Case (b) (hard lumen), the pattern is more pronounced at the outer radius $r_0$ than at the contact radius $r_m$, whereas in Case (e) (soft lumen) the trend of the magnitude is reversed. 
The model thus  captures well the observation that when the contractility of the organoid is disrupted, the apical surface (i.e. the interface between the lumen and the cortex) corrugates more than the basal one (outer surface of the organoid). 
\\
\indent 
Moreover, our model predicts a linear scaling law between the critical wavelength and the initial thickness of the disk. 
When the shear modulus of the core is approximately halved we find that the slope of the thickness/wavelength curve increases by a factor $1.57$ (Fig.~\ref{fig:LIS1}(b)), which is consistent with the one measured value for LIS1 mutant organoids (Fig.~\ref{fig:LIS1}(a)). 
LIS1 mutation is associated with \textit{lissencephaly}, a brain malformation which results in the development a smooth cortex surface. 
In \cite{karzbrun2018human}, the LIS1 mutant organoid where the core softens by a factor 2.8 develops a pattern with a  wavelength magnified by a factor $2.37$ with respect to the control organoid (see Fig.~\ref{fig:LIS1}(a)). 
The mutation not only softens the lumen of the organoid, but it also slows the inward motion of the nuclei. 
This leads to an accumulation of nuclei at the outer surface of the cortex and introduces a compression along the circumference of the cortical ring. 
In effect, our model predicts that for disks with stiffness ratio $\mu=7$ (mutant), the critical remodeling threshold of the cortex $\gamma_C^{\crt}=0.85$ is higher (i.e. buckling occurs earlier) than for the case $\mu=2.5$ (control) where $\gamma_C^{\crt}=0.76$ (see Fig.~\ref{fig:LIS1}(c)).\\

\indent\textbf{Conclusion --\ }Our model predicts the onset of wrinkling in human brain organoids, associated with the early gyrification in the human brain. Both the contraction of the lumen and the motion of the nuclei of neuron progenitor cells within the cortex play a crucial role in the early stages of brain development. Disruption of any of these mechanisms results in alteration of the final folding pattern and leads to brain malformation. In particular, this model gives new insights on the role of the microstructural remodeling of the cortex in determining the lissencephaly pathology, a mutation associated with a smoother cortex. Our results suggest that a reduced contractility of  the core and a slower motion of nuclei both induce a delay in the onset of the instability and trigger a smoother instability pattern. This model and recent works on brain mechanics  further emphasize the crucial role of physical forces for brain development   and function \cite{gogeho15}.

\section*{Acknowledgements}
 The support for  Alain Goriely by the Engineering and Physical Sciences Research Council of Great Britain under research grant EP/R020205/1 is gratefully acknowledged. The work has received from the European Union's Horizon 2020 Research and Innovation Programme under the Marie Sk\l{}odowska-Curie grant agreement  No.705532 (Valentina Balbi and Michel Destrade).

\bibliography{Organoid.bib}
\bibliographystyle{vancouver}
 
\end{document}